\documentclass[twocolumn,amsmath,amssymb,superscriptaddress,aps,prl,a4paper,floatfix]{revtex4-1}
\usepackage{graphicx}
\usepackage[paperwidth=212mm,paperheight=297mm,centering,hmargin=1.65cm,vmargin=2.6cm]{geometry}
\usepackage{color}
\usepackage{dcolumn}
\usepackage{bm}
\usepackage{physics}
\usepackage{float}
\usepackage[separate-uncertainty = true,multi-part-units=single]{siunitx}
\DeclareSIUnit{\molar}{M}
\DeclareSIUnit{\counts}{cnt}
\usepackage{chemformula}
\usepackage{natbib}
\begin{document}

\title{Photon indistinguishability measurements under pulsed and continuous excitation}

\author{Ross~C.~Schofield}
\affiliation{Centre for Cold Matter, Blackett Laboratory, Imperial College London, Prince Consort Road, SW7 2AZ, London, United Kingdom}
\author{Chloe~Clear}
\affiliation{Quantum Engineering Technology Labs, H. H. Wills Physics Laboratory and Department of Electrical and Electronic Engineering, University of Bristol, BS8 1FD, United Kingdom}
\author{Rowan~A.~Hoggarth}
\affiliation{Centre for Cold Matter, Blackett Laboratory, Imperial College London, Prince Consort Road, SW7 2AZ, London, United Kingdom}
\author{Kyle~D.~Major}
\affiliation{Centre for Cold Matter, Blackett Laboratory, Imperial College London, Prince Consort Road, SW7 2AZ, London, United Kingdom}
\author{Dara~P.~S.~McCutcheon}
\affiliation{Quantum Engineering Technology Labs, H. H. Wills Physics Laboratory and Department of Electrical and Electronic Engineering, 
University of Bristol, BS8 1FD, United Kingdom}
\author{Alex~S.~Clark}
\email{alex.clark@imperial.ac.uk}
\affiliation{Centre for Cold Matter, Blackett Laboratory, Imperial College London, Prince Consort Road, SW7 2AZ, London, United Kingdom}
\date{\today}
\begin{abstract}
The indistinguishability of successively generated photons from a single quantum emitter is most commonly measured using two-photon interference at a beam splitter. Whilst for sources excited in the pulsed regime the measured bunching of photons reflects the full wavepacket indistinguishability of the emitted photons, for continuous wave (cw) excitation the inevitable dependence on detector timing resolution and driving strength obscures the underlying photon interference process. Here we derive a method to extract the photon indistinguishability from cw measurements by considering the relevant correlation functions. The equivalence of both methods is experimentally verified through comparison of cw and pulsed excitation of an archetypal source of photons, a single molecule. 
\end{abstract}

\maketitle
\noindent
Many photonic quantum technologies rely on the quantum interference of photons, including linear optical quantum information processing~\cite{Knill2001}, cluster state generation~\cite{Bell2012}, boson sampling~\cite{Bentivegnae2015}, quantum metrology~\cite{Giovannetti2011}, and Bell-state measurements in quantum communication~\cite{Riedmatten2004} and teleportation schemes~\cite{Llewellyn2020}. However, this quantum interference is only possible if the photons used are quantum mechanically indistinguishable, and it is therefore paramount when developing a single photon source that the indistinguishability of emitted photons is quantified. While this can in principle be inferred through separate characterisation of the photons' polarization, spatial, temporal, and frequency modes,  a more rigorous method which directly proves their usefulness is to measure the two-photon interference effect itself.

This two-photon interference effect was first shown by Hong, Ou and Mandel using photons probabilistically generated through spontaneous parametric down-conversion of a pump laser in a nonlinear crystal \cite{Hong1987}. 
Since then routes toward generating photons on-demand have emerged~\cite{Eisaman2011a}, for example those using single quantum emitters such as atoms~\cite{Hijlkema2007}, quantum dots~\cite{Santori2002,Somaschi2016,Wang2019a,Iles-Smith2017}, crystalline defects~\cite{Aharonovich2011} and single molecules~\cite{Toninelli2021}. For these systems it is common to interfere successively emitted photons from a single source by introducing an appropriate delay and mixing the two signals on a beam splitter. The interference is then manifest as a reduction in coincidence counts at the beam splitter outputs, as measured by the second-order correlation function $g^{(2)}(\tau)$.

If the emitter is excited regularly with a pulsed laser, then
the normalised time-integrated difference between $g^{(2)}(\tau)$ measurements for photons input with perpendicular and parallel polarization directly gives the full photon wavepacket indistinguishability $\mathcal{I}=\langle\psi_1|\psi_2\rangle$, where $|\psi_{1,2}\rangle$ represent the quantum states, or wavefunctions, of the two interfering photons at the point of measurement. This reflects the underlying modal purity of the photons and gives the probability of two-photon interference, sometimes called the coalescence probability~\cite{Santori2002,Tomm2021}. 
On the other hand, source excitation with a continuous wave (cw) laser is also commonly used \cite{Patel2008,Trebbia2010,Rezai2018,Lombardi2021}, and the time-resolved $g^{(2)}(\tau)$ credited with indicating the extent of the two-photon interference phenomenon. However, such measurements
are highly dependent on detector timing resolution; the value of $g^{(2)}(0)$ tends to zero for perfect detector resolution regardless of the photon spectral purity~\cite{Iles-Smith2017PRB}, as the measurement is itself effectively a frequency filter. 
While methods to extract detector resolution independent metrics from cw measurements have been proposed~\cite{Proux2015}, they do not give the unitless indistinguishability measure as found in the pulsed case.

In this paper we derive correlation functions for both pulsed and cw excitation of a single photon emitter and develop a method to determine the full photon wavepacket indistinguishability under cw excitation, taking into account the dependence of the measurement on driving strength.
This is experimentally verified through measurements of a single dibenzoterrylene (DBT) molecule in an anthracene host matrix. Pulsed and cw measurements are performed on the same molecule to independently determine $\mathcal{I}$, showing the correspondence of the two methods. This equivalence provides a useful analysis tool for developing on-demand photon sources from single quantum emitters. 

To begin we consider a beam splitter with successively generated photons from a single quantum emitter at its inputs. The emitter is
modelled as a two-level system 
and described by dipole operator $\sigma=\ket{g}\bra{e}$, with $\ket{e}$ and $\ket{g}$ the excited and ground states. Under pulsed excitation the unnormalised second-order correlation function describing photon detection coincidences at the outputs for photons with parallel polarization at the inputs is 
\begin{multline}\label{eq:G2pulpara}
\!\!\!G^{(2)}_{\parallel}(\tau)=\!\!\int^{\infty}_{0}\!\!\!\!dt \big[P_e(t)P_e(t+\tau)-|g^{(1)}(t+\tau,t)|^2\big] ,
\end{multline}
where the first order-correlation function is $g^{(1)}(t_1,t_2)=\expval{\sigma^{\dagger}(t_1)\sigma(t_2)}$, 
the excited state population at time $t$ is 
$P_e(t)=\langle \sigma_{ee} (t) \rangle$ with $\sigma_{ee}=\sigma^{\dagger}\sigma$, 
and the integration over $t$ gives a coincidence probability per pulse~\cite{Kiraz2004}. 
For perpendicular input polarizations 
photon distinguishability is imposed, and the coincidence probability becomes
$
G^{(2)}_{\perp}(\tau)=\int^{\infty}_{0}dt P_e(t)P_e(t+\tau) \, .
$
A detailed derivation can be found in the Supplemental Material.
The photon indistinguishability
is defined as the normalised difference in coincidence events for parallel and perpendicular input polarisations, integrated over all detection time differences $\tau$~\cite{Bylander2003}:
\begin{equation}\label{eq:Ipulg2}
\mathcal{I}=\frac{\int d\tau ~G^{(2)}_{\perp}(\tau)-\int d\tau ~G^{(2)}_{\parallel}(\tau)}{\int d\tau ~G^{(2)}_{\perp}(\tau)} \, .
\end{equation}
For the case of a quantum emitter with spontaneous decay rate $\Gamma_1$ and dephasing rate $\Gamma_2=\Gamma_1/2+\gamma$ where $\gamma$ represents excess dephasing, we find $\mathcal{I}=\Gamma_1/(2\Gamma_2)$.

Under cw excitation conditions, in the steady state the measured coincidences for parallel inputs is found to be 
\begin{equation}
\label{g2parallel}
g_{||}^{(2)}(\tau)
=\frac{1}{2}
+\lim_{t\to\infty}\frac{ g^{(2)}(t,t+\tau)-|g^{(1)}(t+\tau,t)|^2}{2 P_e^2} \, ,
\end{equation}
which in this case is normalised by the square of the excited steady-state population $P_e=\lim_{t\to\infty}P_e(t)$ and we have defined $g^{(2)}(t_1,t_2)=\expval{\sigma^{\dagger}(t_1)\sigma^{\dagger}(t_2)\sigma(t_2)\sigma(t_1)}$~\cite{Unsleber2015}.
For the case of perpendicular inputs where the fields can be treated as uncorrelated we have
\begin{equation}
g_{\perp}^{(2)}(\tau)
=\frac{1}{2}+\lim_{t\to\infty}\frac{ g^{(2)}(t,t+\tau)}{2 P_e^2} \, .
\end{equation}
It is common to consider a reduction in $g^{(2)}_\parallel(\tau)$ at $\tau=0$ as an indication of the probability of two-photon interference and photon purity. However, since $\sigma(t)^2=0$, it follows that $g^{(2)}(t,t)=0$, while $g^{(1)}(t,t)=P_e(t)$, and one can therefore see from Eq.~({\ref{g2parallel}}) that $\smash{g^{(2)}_{||}(0)=0}$ regardless of the photon coherence.
In experiments deviations from this value arise due to detector imperfections being unable to precisely resolve $\tau=0$. As such the value of $g^{(2)}_{||}(0)$ at best reflects a combination of the detector response and photon distinguishability. 
We could, perhaps, integrate over $\tau$ as in the pulsed case, but as these cw quantities give coincidences per unit time and the system is driven, the time-integrals diverge. 
To overcome this, we propose to first subtract the excited steady-state population which recovers a convergent integral similar to that in Eq.~(\ref{eq:Ipulg2}), which after cancellations becomes
\begin{equation}
\tilde{\mathcal{I}}(S)=
\frac{\int d\tau[1-g^{(2)}_{\parallel}(\tau)]-\int d\tau[1-g^{(2)}_{\perp}(\tau)]}{\int d\tau [1-g^{(2)}_{\perp}(\tau)]} \, ,
\label{eq:Ig2CW}
\end{equation}
which in general is a function of the cw driving strength described by the saturation parameter $S$. Our crucial observation is that in the limit of weak driving $\tilde{\mathcal{I}}(0)=\mathcal{I}$, and we see that cw measurement contains the true photon indistinguishability that we seek.

It is not, of course, possible to measure the correlation function at $S=0$ as no photons are emitted. We therefore seek an analytical expression for $\tilde{I}(S)$, from which $\mathcal{I}$ can be extracted. To do so we consider an incoherently driven effective two-level system, obtained
by adiabatic elimination of the fast decaying higher energy state used for off-resonant excitation. This is valid provided decay from the pump level at a rate $\beta$ is fast compared to the other system rates ($\beta\gg S\Gamma_1, \Gamma_2$). See Supplemental Material for details.

The result is a second order Born-Markov master equation for the effective two-level system density operator $\rho$:
\begin{equation}
\partial_t \rho(t) = \Gamma_1(\mathcal{L}_{\sigma}[\rho(t)]+ S\mathcal{L}_{\sigma^{\dagger}}[\rho(t)])+2
\gamma \mathcal{L}_{\sigma_{ee}}[\rho(t)] \, ,
\label{eq:2LS_ME}
\end{equation}
where $\mathcal{L}_{X}[\rho(t)]=X\rho(t)X^{\dagger}- \frac{1}{2}\qty\big{ X^{\dagger}X,\rho(t)}$ is a Lindblad operator. The incoherent driving is captured by the term involving 
$S=\Omega^2/(\beta\Gamma_1)$ 
where $\Omega$ is the Rabi frequency between the ground and higher energy pump level. Using this master equation and the quantum regression theorem~\cite{Guarnieri2014} we find the correlation functions 
are
\begin{equation}
    g^{(2)}_{\parallel}(\tau)=1-\frac{\mathcal{V}}{2}e^{-\Gamma_1(1+S)|\tau|}\Big(1+\mathcal{M}\,e^{-2\gamma|\tau|}\Big),
    \label{eq:cwdip}
\end{equation}
where we have introduced $\mathcal{V}$ to account for any imperfection in anti-bunching visibility and $\mathcal{M}$ to account for any modal distinguishability with no temporal dependence, such as incoherent sideband emission or polarization mismatch. For perpendicular polarization $g^{(2)}_{\perp}(\tau)$ is given by Eq.~(\ref{eq:cwdip}) with $\mathcal{M}=0$.
Using these in Eq.~(\ref{eq:Ig2CW}) we find
\begin{equation}
\tilde{\mathcal{I}}(S)=\mathcal{M}\frac{\Gamma_1(1+S)}{\Gamma_1(1+S)+2\gamma} \, ,
\label{eq:Icw}
\end{equation}
which allows for cw measurements of $g^{(2)}_{\parallel/\perp}(\tau)$ to be integrated at a known $S$ and extrapolated to $S=0$ to give $\mathcal{I}$. 
The effective two-level system model from which Eq.~(\ref{eq:Icw}) is derived holds for $\beta\gg S \Gamma_1$, which is well within the validity of our system parameters (see Supplemental Material). 
At high $S$ 
stimulated emission from $S_{1,n>0}$ leads to deviations from the behaviour described in Eq.~({\ref{eq:2LS_ME}})~\cite{Schofield2018}. 

\begin{figure}
    \centering
    \includegraphics[width=0.9\columnwidth]{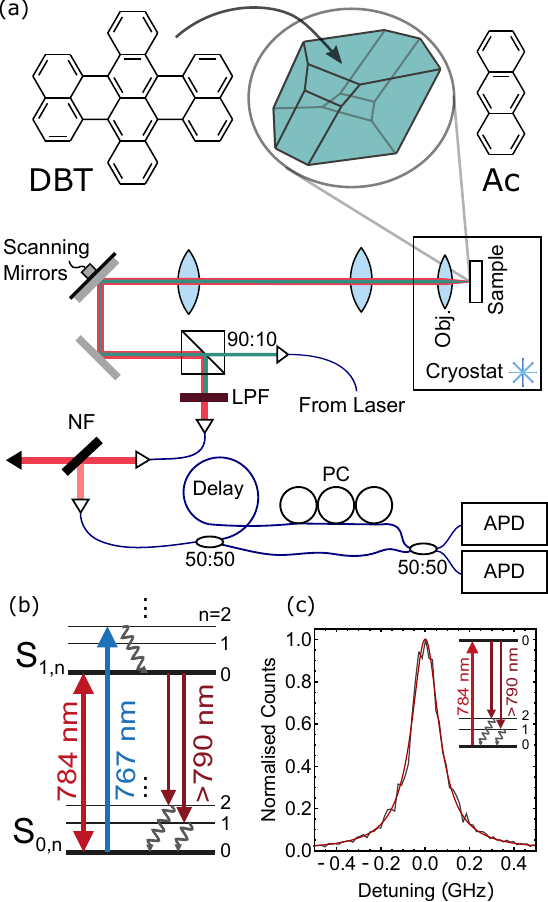}
    \caption{(a) Schematic of the nanocrystal and experimental apparatus used. Diagrams of dibenzoterrylene (DBT), anthracene (Ac) and a DBT containing Ac nanocrystal are shown. In the simplified experimental apparatus, blue lines are single mode fibers, green is excitation laser, dark red is all molecular emission and light red is ZPL emission. The confocal microscope setup consists of; 90:10: 90\% reflection, 10\% transmission beam splitter; Obj.: Objective lens; LPF: long-pass filter; NF: notch reflection filter. The interferometer consists of; 50:50: 50\% reflection, 50\% transmission single mode fiber beam splitter; PC: fiber polarization controller; APD: avalanche photodiode. (b) Energy level diagram of the DBT molecule used showing the two electronic energy levels and the associated vibrational levels. Approximate wavelengths of the transitions are shown. (c) Fluorescence excitation spectrum of the DBT molecule ZPL showing the change in detected photon counts as the laser is tuned relative to the $S_{0,0}\to S_{1,0}$ transition frequency of the molecule. Data (grey) are fit with a Lorentzian (red) to obtain the linewidth. Inset: Energy levels showing the excitation (\SI{784}{\nano\meter}) and collected fluorescence ($>\SI{790}{\nano\meter}$) wavelengths.}
    \label{fig:fig1}
\end{figure}

We now turn to indistinguishability measurements of photons emitted by a single DBT molecule to verify our theory. To isolate a single molecule we used a DBT-doped anthracene nanocrystal grown using a re-precipitation technique \cite{Pazzagli2018}, see Fig.~\ref{fig:fig1}(a). This crystal was deposited onto a gold-coated silicon substrate with a \SI{85}{\nano\meter} silica spacer layer and protected with a \SI{200}{\nano\meter} thick layer of PVA. The gold mirror increases the collection efficiency of light from the molecule~\cite{Clear2020}. The sample was cooled to \SI{4.7}{\kelvin} in a closed-cycle cryostat (Montana Cryostation) that forms part of a confocal microscope shown in Fig.~\ref{fig:fig1}(a). A nanocrystal was selected and illuminated with a cw Ti:Sapphire laser (MSquared, SolsTiS), directed using the scanning mirrors. Fig.~\ref{fig:fig1}(b) shows the energy level diagram of a DBT molecule and the laser frequencies used for excitation. The laser was tuned in frequency to identify a molecule resonance through excitation of the $S_{0,0}\to S_{1,0}$ zero-phonon line (ZPL) transition, around 784\,nm, while  the red-shifted fluorescence ($>$790\,nm), shown in Fig.~\ref{fig:fig1}(c), from the $S_{1,0}\to S_{0,n>0}$ transitions was collected in a multi-mode fiber and detected with a silicon avalanche photodiode (APD). 

A single DBT resonance was found at \SI{784.45}{\nano\meter}, and initial characterisation was performed by repeating scans at increasing excitation powers to determine the maximum count rate and linewidth $\Delta\nu$ of the molecule at each power. This was used to determine the dephasing rate $\Gamma_2$ and saturation behaviour of the molecule using the power-broadening relationship $\Delta\nu=\Gamma_2/(\pi)\sqrt{1+S}$~\cite{Grandi2016}.
From this we find $\Gamma_2=2\pi \times \SI{35(4)}{\mega\hertz}$.  
\begin{figure*}[t!]
    \includegraphics[width=2\columnwidth]{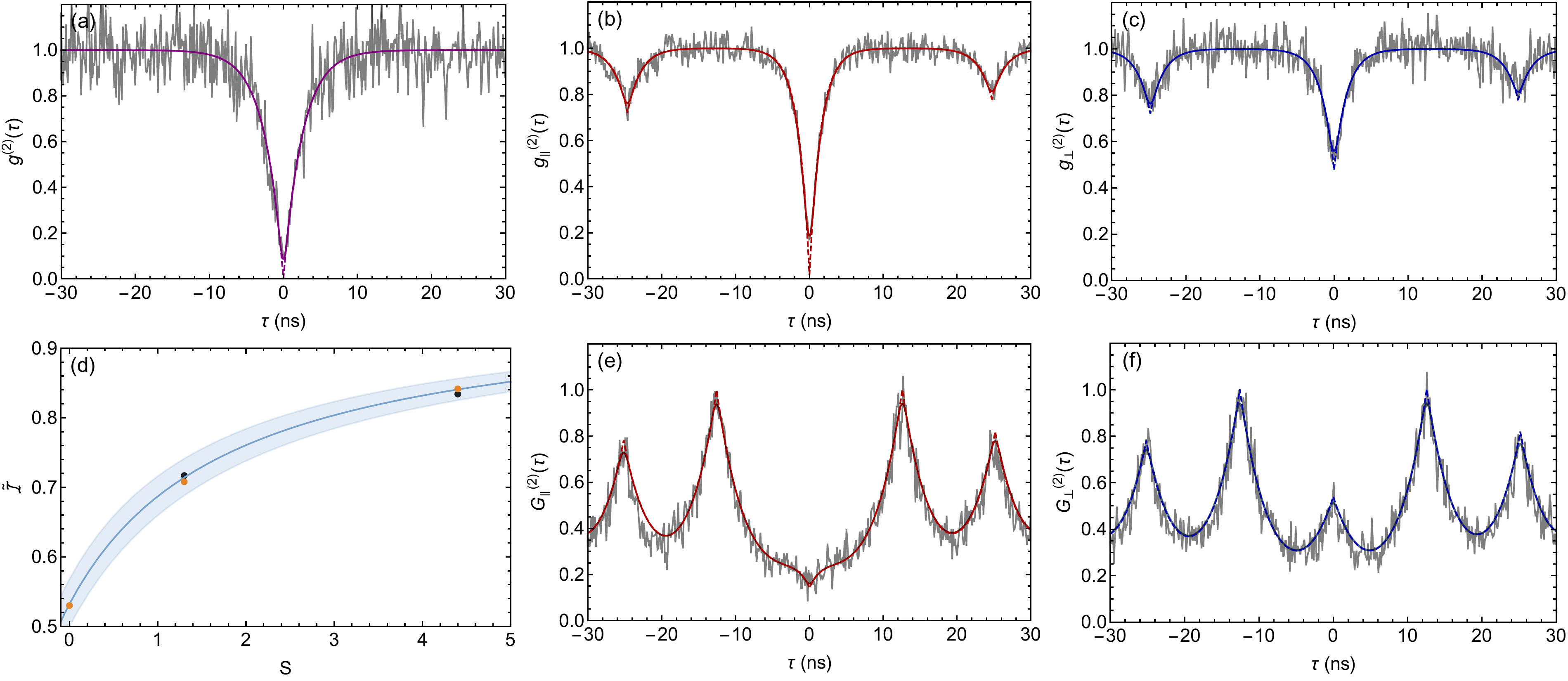}
    \caption{Continuous wave (cw) and pulsed measurements of photon indistinguishability performed on the same molecule. Measurements are in black, and colored curves show theoretical fits with (solid) and without (dashed) accounting for detector response. (a) A cw intensity correlation $\smash{g^{(2)}(\tau)}$ with theoretical fits using Eq.~(\ref{eq:g2}). (b) A cw $\smash{g^{(2)}_{\parallel}(\tau)}$ measurement with theoretical curves using  Eq.~(\ref{eq:cwdip}). (c) A cw $g^{(2)}_{\perp}(\tau)$ measurement with theroetical curves using Eq.~(\ref{eq:cwdip}). (d) Extracted $\mathcal{\tilde{I}}$ as a function of saturation parameter ($S$). The prediction from Eq.~(\ref{eq:Icw}) is shown as the solid line, with the shaded region indicating uncertainties in $\Gamma_1$ and $\Gamma_2$. Data points are from Eq.~(\ref{eq:Icw}) using integration of the data (black) and fitted functions (orange). The data point at $S=0$ is from pulsed measurements in parts (e) and (f). (e) A pulsed excitation $G^{(2)}_{\parallel}(\tau)$ measurement displaying anti-bunching and two-photon interference, with theory curves using Eq.~(\ref{eq:cwdip}), modified to account for the pulsed behaviour. (f) A pulsed excitation $G^{(2)}_{\perp}(\tau)$ measurement.}
    \label{fig:fig2}
\end{figure*}
The cw laser was then tuned to \SI{766.67}{\nano\meter} resonant with a $S_{0,0}\to S_{1,n>0}$ transition, shown as a blue arrow in Fig.~\ref{fig:fig1}(b). The collection was changed to use a single mode fiber and a narrowband (0.15\,nm) tunable reflective notch filter positioned before the APDs. 
The filter response function and the expected effect on the molecule spectrum is shown in the Supplemental Material. Only the coherent emission from the $S_{1,0}\to S_{0,0}$ ZPL transition will provide measurable interference; the narrowband filter is used to remove emission from the phonon sideband \cite{Clear2020} and $S_{1,0}\to S_{0,n>0}$ transitions.
After filtering we expect a ratio of coherent to total emission of $>\SI{99}{\percent}$. 

To verify single photon emission a Hanbury Brown and Twiss 
$g^{(2)}(\tau)$ measurement was performed, shown in Fig.~\ref{fig:fig2}(a), by splitting the fluorescence directly on a 50:50 beam splitter before two APDs. Fitting the data using~\cite{Grandi2016}
\begin{equation}
    g^{(2)}(\tau)=1-\mathcal{V}\,e^{-\Gamma_1(1+S)|\tau|} \, ,
    \label{eq:g2}
\end{equation}
we find a visibility of $\mathcal{V}=0.98^{+0.02}_{-0.03}$, which when accounting for detector timing jitter gives $\mathcal{V}=1.00^{+0}_{-0.03}$ indicating we are observing a single emitter. Accounting for the measured saturation parameter $S$ (see Supplemental Material) we find a population decay rate of $\Gamma_1=2\pi \times \SI{40(2)}{\mega\hertz}$. This is independently verified using a time-correlated single photon counting measurement with a pulsed Ti:Sapphire laser (Coherent, Tsunami) which gives $\Gamma_1=2\pi \times \SI{39(3)}{\mega\hertz}$. Comparison of the dephasing and population decay rates gives $\Gamma_1/2\Gamma_2=0.57\pm0.09$, typical at these temperatures due to the excess thermal dephasing \cite{Clear2020,Grandi2016}.

Turning now to measuring indistinguishability using cw two-photon interference, the fluorescence was sent to the fiber-based interferometer shown in Fig.~\ref{fig:fig1}(a). A 50:50 fiber beam splitter and delay fiber was used to temporally overlap
photons at a second beam splitter, where two-photon interference occurs. A fiber polarisation controller allowed for measurements of photons with parallel or perpendicular polarization. 
The results of the parallel and perpendicular 
interference measurements at $S=1.3\pm0.1$ are shown in Fig.~\ref{fig:fig2}(b) and (c). The data falls below $0.5$ 
in the parallel case due to photon interference and coalescence. The side dips arise from anti-bunching at different time delays due to the different combinations of possible optical paths~\cite{Rezai2018}. Fitting these side dips determines the $S$ and $\mathcal{V}$ parameters. Equation~(\ref{eq:cwdip}), convolved with the detector response function, is plotted over the data using the determined experimental parameters and $\mathcal{M}=0.98$, showing a good correspondence between the measurement and expected result. The non-convolved function is shown as a dashed line. 
This is repeated for the orthogonal polarization $g_{\perp}^{(2)}(\tau)$ measurement, shown in Fig.~\ref{fig:fig2}(c), where $\mathcal{M}=0.04$. For a measurement with perfectly orthogonal polarization $\mathcal{M}=0$, however polarization drift during measurement resulted in a small two-photon interference contribution. This is characterised in the Supplementary Material.

Figure~\ref{fig:fig2}(d) shows the ratio of the integrals described in Eq.~(\ref{eq:Ig2CW}) for measurements taken at $S=1.3\pm0.1$ and $4.4\pm0.2$; both values are well within the validity range for our model.
Values of $\tilde{\mathcal{I}}$ based on the raw data (black) and the de-convolved functions (orange) are shown, and agree within error. Fitting Eq.~(\ref{eq:Icw}) with $\mathcal{M}$ as the free variable gives $\mathcal{M}=0.96\pm0.01$ and an indistinguishability of $\mathcal{I}=0.53\pm0.01$ at $S = 0$. 

To confirm this result, we now turn to using pulsed excitation. We use  
a pulsed Ti:Sapphire laser (Spectra-Physics, Tsunami) tuned to 766\,nm and filtered to a bandwidth of 5\,nm to excite the molecule again on a $S_{0,0}\to S_{1,n>0}$ transition. The parallel and perpendicular correlation functions are shown in Figs.~\ref{fig:fig2}(e) and (f), 
with each normalised to one. Here the $\sim12.5$\,ns laser repetition period is only a few times longer than the $\sim4$\,ns lifetime of the molecule, and as such photons from subsequent pulses partially overlap. When taking the difference between the $G^{(2)}_{\parallel/\perp}(\tau)$ measurements in Eq.~(\ref{eq:Ipulg2}) contributions from the overlapping side features cancel, though this is not the case in
the denominator.  
This requires fitting to subtract the contribution of side features from the data to give the true integral of the central feature needed to quantify the indistinguishability according to Eq.~(\ref{eq:Ipulg2}). In doing so we find $\mathcal{I}=0.48\pm0.02$. This is lower than in the cw measurement due 
to imperfect temporal overlap arising from a mismatch of the fiber delay and the pulse repetition period in our interferometer. This can be accounted for with a correction factor of $e^{-\Gamma_1\Delta\tau}$ where $\Delta\tau$ is the time difference between the laser repetition period and the delay time from the fiber \cite{Schneiderbook2015}. This is $0.91\pm0.02$ for our setup, and after this correction we find $\mathcal{I}=0.53\pm0.02$, matching the value found through cw excitation. This is in line with the expected $\mathcal{I}$ value when considering $\mathcal{I}=\Gamma_1/2\Gamma_2 \times \mathcal{M}=0.54\pm0.09$, where the polarisation drift ($0.95$) and branching ratio (0.99) are contributing to $\mathcal{M}$.
The indistinguishability is limited primarily by excess thermal dephasing, which greater cooling can eliminate \cite{Clear2020,Trebbia2010}. Additionally, these measurements highlight the potential of single molecules for quantum technology applications \cite{Toninelli2021} when considering their integration into nanophotonic structures such as waveguides \cite{Boissier2021,Rattenbacher2019}, patterned polymers \cite{Ciancico2019, Schofield2020} and cavities \cite{Wang2019}.

In this work we have shown a method to extract the full wavepacket indistinguishability of photons from a cw-excited single quantum emitter using two-photon interference measurements.  
This was experimentally verified by comparing photon indistinguishability found from cw and pulsed measurements performed on a single DBT molecule at cryogenic temperatures. Previous discussion of cw two-photon interference measurements has been limited to stating $g^{(2)}_{\parallel/\perp}(0)$ values, a metric that 
is not independent of the detector timing resolution. 
We note that our underlying theoretical treatment holds for more complex systems. We already account for the coherent excitation to a third energy level and find a suitable parameter range for disregarding coherent effects, and could be expanded to considering the effects of optical cavities on photon emission~\cite{Iles-Smith2017}.
The interference of photons from two separate emitters has also been demonstrated with defects in diamond~\cite{Sipahigil2014,Bernien2012}, quantum dots~\cite{Flagg2010,Patel2010} and molecules \cite{Lettow2010}. Our method could be straightforwardly extended to account for the effects of driving on these systems, and could include further parameters such as different central frequencies and dephasing rates of the two emitters used. 

In contrast to the pulsed case, determining indistinguishability from cw excitation requires multiple measurements at known pump powers, or a single measurement at a known $S$.
However, it allows extraction of the indistinguishability from the raw data, independent of the ratio of emitter lifetime to laser repetition rate. There is also no requirement for the interferometer delay to be a multiple of the laser repetition period, and cw excitation may also be more convenient due to the higher count rates and the greater spectral selectivity provided. 
These advantages open the possibility of performing multimode quantum interference experiments such as boson sampling~\cite{Bentivegnae2015} with a single cw-driven emitter and appropriate optical delay lines, thereby simplifying experimental demonstrations. 

\begin{acknowledgements}
We thank Jon Dyne and Dave Pitman for their expert mechanical workshop support. This work was supported by EPSRC (EP/P030130/1, EP/P01058X/1, EP/R044031/1, EP/S023607/1, and EP/L015544/1), the Royal Society (UF160475), and the EraNET Cofund Initiative QuantERA under the European Union’s Horizon 2020 research and innovation programme, Grant No. 731473 (ORQUID Project). 
\end{acknowledgements}

\bibliography{library}
\end{document}


\title{Supplemental Material:\\Photon indistinguishability measurements under pulsed and continuous excitation}
\author{Ross~C.~Schofield}
\affiliation{Centre for Cold Matter, Blackett Laboratory, Imperial College London, Prince Consort Road, SW7 2AZ, London, United Kingdom}
\author{Chloe~Clear}
\affiliation{Quantum Engineering Technology Labs, H. H. Wills Physics Laboratory and Department of Electrical and Electronic Engineering, University of Bristol, BS8 1FD, United Kingdom}
\author{Rowan~A.~Hoggarth}
\affiliation{Centre for Cold Matter, Blackett Laboratory, Imperial College London, Prince Consort Road, SW7 2AZ, London, United Kingdom}
\author{Kyle~D.~Major}
\affiliation{Centre for Cold Matter, Blackett Laboratory, Imperial College London, Prince Consort Road, SW7 2AZ, London, United Kingdom}
\author{Dara~P.~S.~McCutcheon}
\affiliation{Quantum Engineering Technology Labs, H. H. Wills Physics Laboratory and Department of Electrical and Electronic Engineering, 
University of Bristol, BS8 1FD, United Kingdom}
\author{Alex~S.~Clark}
\email{alex.clark@imperial.ac.uk}
\affiliation{Centre for Cold Matter, Blackett Laboratory, Imperial College London, Prince Consort Road, SW7 2AZ, London, United Kingdom}
\begin{abstract}
In this supplement we first derive the indistinguishability of an emitter measured using two-photon interference with a Hong--Ou--Mandel (HOM) interferometer \cite{HOM1987} for non-resonant pulsed excitation. This formalism is then extended for continuous wave (cw) excitation considering a coherently driven three-level system. To find an analytical form for extracting indistinguishability, we simplify the system by the adiabatic elimination of the fast decaying higher energy state used for off-resonant excitation. This formalism is extended for the case of differing driving strengths between perpendicular and parallel polarisation alignment of the interferometer. Moreover, the effect of the phonon sideband is accounted for. In the latter half of the supplement we present experimental details. Measurements performed to characterise the DBT molecule used in the main paper are presented. The effect of filtering on the ratio of coherent to total emission is discussed. The experimental parameters of the interferometer are discussed and an equation that accounts for these is presented. Finally, the indistinguishability measurements performed at higher driving strength are shown. 
\end{abstract}
\maketitle

\section{Interference theory}
We seek to derive the general second-order correlation function for the output fields of a two-photon interference experiment. For this set up we
have two (positive) input fields $E^{(+)}_{1}(t)$ and $E^{(+)}_{2}(t)$ which pass through a 50:50 beam splitter and are related to the (positive) detected fields $E^{(+)}_{3}(t)$ and $E^{(+)}_{4}(t)$ by $E^{(+)}_{3}(t)=\frac{1}{\sqrt{2}}(E^{(+)}_{1}(t)+E^{(+)}_{2}(t))$ and $E^{(+)}_{4}(t)=\frac{1}{\sqrt{2}}(E^{(+)}_{2}(t)-E^{(+)}_{1}(t))$ \cite{Kiraz2004}.
The unnormalised general cross-correlation function for the output fields with parallel polarisation between interferometer arms is
\begin{equation}\label{eq:g2OUT}
G^{(2)}_{\parallel}(\tau,t)=\expval{E^{(-)}_{3}(t)E^{(-)}_{4}(t+\tau)E^{(+)}_{4}(t+\tau)E^{(+)}_{3}(t)},
\end{equation}
where the output field $E_3$ is detected at $t$ and the output field $E_4$ is detected at $t+\tau$ leading us to define $\tau$ as the time delay between the two detection measurements.
Substituting the input fields into Eq. \ref{eq:g2OUT} we find
\begin{multline}
\label{eq:g2IN}
G^{(2)}_{\parallel}(\tau,t)=\frac{1}{4}\Big\langle\big(E^{(-)}_{1}(t)+E^{(-)}_{2}(t)\big)\big(E^{(-)}_{2}(t+\tau)-E^{(-)}_{1}(t+\tau)\big)\\  \big(E^{(+)}_{2}(t+\tau)-E^{(+)}_{1}(t+\tau)\big)\big(E^{(+)}_{1}(t)+E^{(+)}_{2}(t)\big)\Big\rangle.
\end{multline}
Simplifying Eq. \ref{eq:g2IN} as we assume $E_{1}^{(+)}$ and $E_{2}^{(+)}$ originate from the same emitter and are statistically independent; we therefore factorise and drop the numbered subscript. Expanding the correlation function in Eq. \ref{eq:g2IN} 
gives eight terms which are linear in $\expval{E^{(+/-)}}$ and two terms in the form $\expval{E^{(+/-)}E^{(+/-)}}$ which both go to zero, as expectation values linear in ladder operators are zero \cite{Unsleber2015}. 
We find
\begin{multline}\label{eq:g2INsimp}
G^{(2)}_{\parallel}(\tau,t)=\frac{1}{2}\Big(4G^{(2)}_{HBT}(t,\tau)-\abs\Big{ \expval{E^{(-)}(t+\tau)+E^{(+)}(t)}}^2\\+\expval{E^{(-)}(t)E^{(+)}(t)}\expval{E^{(-)}(t+\tau)E^{(+)}(t+\tau)}\Big),
\end{multline} 
where $G^{(2)}_{HBT}(t,\tau)=\frac{1}{4}\expval{E^{(-)}(t)E^{(-)}(t+\tau)E^{(+)}(t+\tau)E^{(+)}(t)}$ is the Hanbury Brown and Twiss second-order correlation function, relating to the case whereby only one input field is incident on the beam splitter.  

\subsection{Pulsed excitation}
For the non-resonant pulsed excitation of a quantum emitter we can model a two-level system initially populated in its excited state, with spontaneous decay rate $\Gamma_1$ and excess pure dephasing $\gamma$. The dynamics can be described with the second-order Born-Markov master equation
\begin{equation}
\partial_t \rho(t) = \Gamma_1\mathcal{L}_{\sigma}[\rho(t)]+2
\gamma \mathcal{L}_{\sigma^{\dagger}\sigma}[\rho(t)],
\end{equation}
where $\rho(t)$ is the time dependent density operator, $\sigma=\ket{g}\bra{e}$ is the dipole operator with $\ket{e}=(1,0)$ and $\ket{g}=(0,1)$ and $\mathcal{L}_{X}[\rho(t)]=X\rho(t)X^{\dagger}- \frac{1}{2}\qty\big{ X^{\dagger}X,\rho(t)}$ is the Lindblad operator. Setting the input fields as single photons emitted from this quantum emitter, in the far field limit we can set $E^{(+)}(t)$ and $E^{(-)}(t)$ to the dipole operators $\sigma(t)$ and $\sigma^{\dagger}(t)$, where we have dropped numerical factors for clarity. 

To model pulsed excitation of this emitter we change variables to the dipole operators and integrate over $t$ to find the unnormalised ensemble average of coincidence events as
\begin{equation}\label{eq:G2pulpara}
G^{(2)}_{\parallel_{PUL}}(\tau)=\frac{1}{2}\int^{\infty}_{0}d t \Big(\expval{\sigma^{\dagger}(t)\sigma(t)}\expval{\sigma^{\dagger}(t+\tau)\sigma(t+\tau)}-\abs\Big{ \expval{\sigma^{\dagger}(t+\tau)+\sigma(t)}}^2\Big),
\end{equation}
where $G^{(2)}_{HBT}(t,\tau)=0$ under pulsed excitation for a single photon emitter as $\sigma^2=0$ \cite{Unsleber2015}. To find the normalised second order correlation function, we divide by the uncorrelated peak area $A=\int^{\infty}_{0}d t\int d\tau\expval{\sigma^{\dagger}(t)\sigma(t)}\expval{\sigma^{\dagger}(t+\tau)\sigma(t+\tau)}$ to find $g^{(2)}_{\parallel_{PUL}}(\tau)=\frac{1}{A}G^{(2)}_{\parallel_{PUL}}(\tau)$.
For the case of perpendicular polarisation between interferometer arms the fields are uncorrelated giving
\begin{equation}\label{eq:G2pulperp}
G^{(2)}_{\perp_{PUL}}(\tau)=\frac{1}{2}\int^{\infty}_{0}d t \Big(\expval{\sigma^{\dagger}(t)\sigma(t)}\expval{\sigma^{\dagger}(t+\tau)\sigma(t+\tau)}\Big).
\end{equation}

We can readily find the indistinguishability of the emitter from these correlation measurements by integrating over $\tau$ to find \cite{Bylander2003}
\begin{equation}\label{eq:Ipulg2}
\mathcal{I}=\frac{\int d\tau ~G^{(2)}_{\parallel}(\tau)-\int d\tau ~G^{(2)}_{\perp}(\tau)}{\int d\tau ~G^{(2)}_{\perp}(\tau)}.
\end{equation}
Substituting in $G^{(2)}_{\parallel_{PUL}}(\tau)$ and $G^{(2)}_{\perp_{PUL}}(\tau)$ into Eq.~\ref{eq:Ipulg2}
we find
\begin{equation}\label{eq:Ipul}
\mathcal{I}=\frac{\int d t\int d\tau\abs\Big{ \expval{\sigma^{\dagger}(t+\tau)+\sigma(t)}}^2}{\int d t\int d\tau\expval{\sigma^{\dagger}(t)\sigma(t)}\expval{\sigma^{\dagger}(t+\tau)\sigma(t+\tau)}}.
\end{equation}
Using quantum regression theorem we can evaluate the correlation function in the numerator of Eq.~\ref{eq:Ipul} to give $|\expval{\sigma^{\dagger}(t+\tau)\sigma(t)}|^2=|\mathrm{Tr}_S[\sigma^{\dagger} e^{\mathcal{L}\tau}\sigma (e^{\mathcal{L}t}\rho_S(0))]|^2=e^{-2\Gamma_1 t}e^{-(\Gamma_1+2\gamma)\tau}$, where $\mathcal{L}$ is the Liouvillian super-operator and the initial population resides in the excited state such that $p_s(0)=(1,0,0,0)$ \cite{QRT}. We evaluate the integrand in the denominator using the same approach, to find $\expval{\sigma^{\dagger}(t)\sigma(t)}\expval{\sigma^{\dagger}(t+\tau)\sigma(t+\tau)}=e^{-\Gamma_1 t}e^{-\Gamma_1 (t+\tau)}$. Performing the integral over both $t$ and $\tau$ we come to the familiar relation 
\begin{equation}\label{eq:Ianalytical}
\mathcal{I}_{\text{pulsed}}=\frac{\Gamma_1}{2\Gamma_2},
\end{equation}
where $\Gamma_2=\frac{\Gamma_1}{2}+\gamma$ is the dephasing rate.

\subsection{Continuous driving}
For the case of non-resonant cw driving the experimentally determined second-order correlation functions take on a different form; instead of measuring the ensemble average over coincidence events we require the steady-state function taking $t\rightarrow \infty$ giving
 \begin{multline}\label{eq:g2CW}
G^{(2)}_{\parallel_{CW}}(\tau)=\lim_{t\to\infty}\frac{1}{2}\Big(4G^{(2)}_{HBT}(t,\tau)-\abs\Big{ \expval{E^{(-)}(t+\tau)+E^{(+)}(t)}}^2\\+\expval{E^{(-)}(t)E^{(+)}(t)}\expval{E^{(-)}(t+\tau)E^{(+)}(t+\tau)}\Big).
\end{multline} 
Substituting in the dipole operator as we assume we are in the far field limit as before and normalising this function with the excited steady state population $\expval{\sigma^{\dagger}\sigma}_{ss}=\lim_{t\to \infty}\expval{\sigma^{\dagger}(t)\sigma(t)}$ (noting this is defined as $P_e$ in the main paper), we find
\begin{equation}\label{eq:g2paracw}
g_{||_{CW}}^{(2)}(\tau)=\frac{1}{2{\expval{\sigma^{\dagger}\sigma}_{ss}^2}}\Big(\lim_{t\to\infty}\big( \expval{\sigma^{\dagger}(t)\sigma^{\dagger}(t+\tau)\sigma(t+\tau)\sigma(t)}-\abs\Big{ \expval{\sigma^{\dagger}(t+\tau)\sigma(t)}}^2\big)+\expval{\sigma^{\dagger}\sigma}_{ss}^2\Big).
\end{equation}
For the case of perpendicular polarisation alignment where the fields are uncorrelated we have
\begin{equation}\label{eq:g2perpcw}
g_{\perp_{CW}}^{(2)}(\tau)=\frac{1}{2\expval{\sigma^{\dagger}\sigma}_{ss}^2}\Big(\lim_{t\to\infty} \expval{\sigma^{\dagger}(t)\sigma^{\dagger}(t+\tau)\sigma(t+\tau)\sigma(t)}+\expval{\sigma^{\dagger}\sigma}_{ss}^2\Big).
\end{equation}
It is evident that we cannot use the same methodology to find the indistinguishability as in Eq. \ref{eq:Ipul} as this gives a divergent result when integrating over $\tau$. However, if we subtract the steady-state population squared (the normalisation factor) we have convergent integrals and can postulate that the indistinguishability can be found from
\begin{equation}\label{eq:ICWgg}
\tilde{\mathcal{I}}(S)=
\frac{\int d\tau(1-g^{(2)}_{\parallel_{CW}}(\tau))-\int d\tau(1-g^{(2)}_{\perp_{CW}}(\tau))}{\int d\tau (1-g^{(2)}_{\perp_{CW}}(\tau))} \, .
\end{equation}
Substituting in the $g^{(2)}_{\parallel_{CW}}(\tau)$ and $g^{(2)}_{\perp_{CW}}(\tau)$ into Eq. \ref{eq:ICWgg} we find
\begin{equation}\label{eq:ICWcorr}
\tilde{\mathcal{I}}(S)=\frac{\int d\tau \lim_{t\to\infty} \abs\Big{ \expval{\sigma^{\dagger}(t+\tau)\sigma(t)}}^2/{\expval{\sigma^{\dagger}\sigma}^2_{ss}} }{\int d\tau~1-\lim_{t\to\infty} \expval{\sigma^{\dagger}(t)\sigma^{\dagger}(t+\tau)\sigma(t+\tau)\sigma(t)}/{\expval{\sigma^{\dagger}\sigma}^2_{ss}}}.
\end{equation}

\subsubsection{Coherent non-resonant driving}
\begin{figure}[h]
\centering
  \includegraphics[scale=0.70]{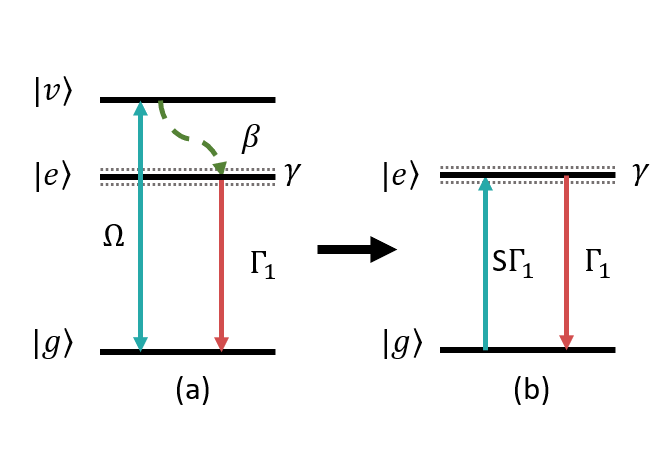}
\caption{(a) Schematic diagram of non-resonant driving from the ground $\ket{g}$ to a higher vibrational level $\ket{v}$, modelled by coherent driving with the Rabi frequency $\Omega$. The fast non-radiative decay rate from $\ket{v}\rightarrow \ket{e}$ is given by $\beta$. Spontaneous emission from the excited state $\ket{e}$ is given by $\Gamma_1$ and pure dephasing is given by $\gamma$. (b) Effective two level system by adiabatic elimination of the pump level, giving a driving rate $S\Gamma_1$ with the saturation parameter S.}
\label{fig:AdiabticElevel}
\end{figure}

To check the validity of this postulated form to find indistinguishability from $\tilde{\mathcal{I}}(S)$ in Eq. \ref{eq:ICWcorr}, a three-level non-resonant driving model shown in Fig. \ref{fig:AdiabticElevel}(a) is considered. Defining the states $\ket{v}=(1,0,0), \ket{e}=(0,1,0),\ket{g}=(0,0,1)$ and the operators $\sigma=\ket{g}\bra{e}$, $\sigma_{vg}=\ket{v}\bra{g}$ and $\sigma_{ev}=\ket{e}\bra{v}$. The subsequent Born-Markov second-order master equation for this system is
\begin{equation}\label{eq:3LS_ME}
\partial_t \rho(t) =-i[H_s,\rho]+ \Gamma_1\mathcal{L}_{\sigma}[\rho(t)]+\beta \mathcal{L}_{\sigma_{ev}}[\rho(t)]+2
\Gamma_1 \mathcal{L}_{\sigma^{\dagger}\sigma}[\rho(t)],
\end{equation}
with $H_s=\Omega/2(\sigma_{vg}+\sigma_{vg}^{\dagger})$ representing the coherent driving with Rabi frequency $\Omega$. For a typical DBT molecule the decay rate from the first localised vibrational mode is  approximately $\beta \approx 2500\times\Gamma_1$ \cite{Clear2020}. The excited steady state population for this system is $\rho_{ee}(\infty)=\expval{\rho(\infty)}{e}=\frac{\frac{\Omega^2}{\beta\Gamma_1}}{1+\frac{\Omega^2}{\beta\Gamma_1}+\frac{2\Omega^2}{\beta^2}}$. We can set saturation to $S=\frac{\Omega^2}{\beta\Gamma_1}$ as $\beta \gg \Gamma_1$ which gives $\rho_{ee}(\infty)=\frac{S}{1+S(1+\frac{2\Gamma_1}{\beta})}\approx \frac{S}{1+S}$.

\begin{figure}[t]
\centering
  \includegraphics[scale=0.60]{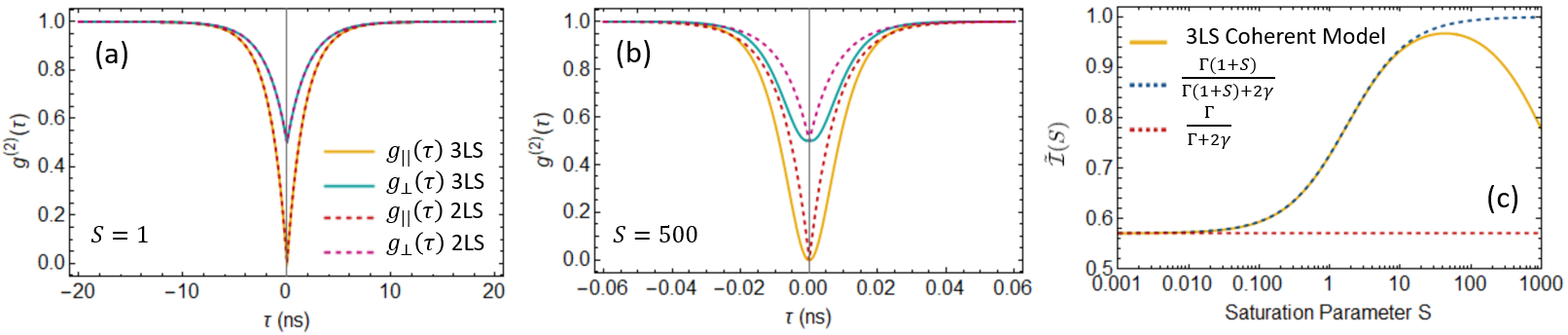}
\caption{(a) and (b) $g^{(2)}_{\perp/\parallel_{CW}}(\tau)$ calculations from the three level system and effective two level system models for varying driving strengths (a) $S=1$ and (b) $S=500$.
(c) Indistinguishability calculated from the three-level system coherent model (solid yellow line). The analytic form of $\tilde{\mathcal{I}}(S)$ from an effective non-resonantly driven two-level system dashed blue. Red line is full photon wavepacket indistinguishability using the parameters of the experimentally measured molecule giving $\mathcal{I}=57\%$.}
\label{fig:Iresults}
\end{figure}

Numerical calculations of $g_{\perp_{CW}}^{(2)}(\tau)$ and $g_{||_{CW}}^{(2)}(\tau)$ for weak $S=0.01$ and strong $S=500$ driving strengths are shown in Fig.~\ref{fig:Iresults}(a) and (b). The parameters for these calculations are the same as presented in the main manuscript with, $\Gamma_1=2\pi \times \SI{40(2)}{\mega\hertz}$ and the dephasing rate $\Gamma_2=2\pi \times \SI{35(4)}{\mega\hertz}$. Numerical simulations of $\tilde{\mathcal{I}}(S)$ calculated from $g_{\perp_{CW}}^{(2)}(\tau)$ and $g_{||_{CW}}^{(2)}(\tau)$ for varying saturation strengths $S$ are shown in Fig.~\ref{fig:Iresults}(c). 

\subsubsection*{Adiabatic elimination of the pump level}

It is advantageous to have an analytical form for $\tilde{\mathcal{I}}(S)$ as this allows for the extraction of indistinguishability from experiment. To do so we derive an effective two-level system by adiabatically eliminating the higher order energy state, see Fig.~\ref{fig:AdiabticElevel}. Starting with the optical Bloch equations for the three level non-resonantly driven system  derived from Eq.~\ref{eq:3LS_ME}, we find 
\begin{equation}\label{eq:RHOvv}
    \dot{\rho}_{vv}(t)=\frac{i\Omega}{2}(\rho_{vg}(t)-\rho_{gv}(t))-\beta\rho_{vv}(t),
\end{equation}
\begin{equation}\label{eq:RHOee}
    \dot{\rho}_{ee}(t)=-\Gamma_1\rho_{ee}(t)+\beta \rho_{vv}(t),
\end{equation}
\begin{equation}\label{eq:RHOgg}
    \dot{\rho}_{gg}(t)=-\frac{i\Omega}{2}(\rho_{vg}(t)-\rho_{gv}(t))+\Gamma_1 \rho_{ee}(t),
\end{equation}
\begin{equation}\label{eq:RHOgv}
    \dot{\rho}_{gv}(t)=\frac{i\Omega}{2}(\rho_{gg}(t)-\rho_{vv}(t))-\frac{\beta}{2} \rho_{gv}(t),
\end{equation}
\begin{equation}\label{eq:RHOge}
    \dot{\rho}_{ge}(t)=-\frac{i\Omega}{2}\rho_{ve}(t)-\frac{\Gamma_1}{2}\rho_{ge}(t)-\gamma\rho_{ge}(t),
\end{equation}
\begin{equation}\label{eq:RHOve}
    \dot{\rho}_{ve}(t)=-\frac{i\Omega}{2}\rho_{ge}(t)-\frac{\Gamma_1}{2}\rho_{ve}(t)-\frac{\beta}{2}\rho_{ve}(t)-\gamma\rho_{ve}(t),
\end{equation}
where $\rho_{XY}(t)=\mel{X}{\rho(t)}{Y}$ \cite{DeValcarcel2006}. 
Solving firstly Eq. \ref{eq:RHOgv} with an integrating factor to find 
\begin{equation}\label{eq:INTgv}
    \rho_{gv}(t)=\frac{i\Omega}{2}\int^t_0d t'e^{-\frac{\beta}{2}(t-t')}(\rho_{gg}(t')-\rho_{vv}(t')),
\end{equation}
which can be solved for the case of $\beta\gg\Omega$ to give $\rho_{gv}(t)\approx\frac{i\Omega}{\beta}(\rho_{gg}(t)-\rho_{vv}(t))$, and by similar methodology $\rho_{vg}(t)\approx-\frac{i\Omega}{\beta}(\rho_{gg}(t)-\rho_{vv}(t))$. Using these forms for $\rho_{gv}(t)$ and $\rho_{vg}(t)$ and substituting into Eq. \ref{eq:RHOvv} we find
\begin{equation}\label{eq:RHOvv2}
    \dot{\rho}_{vv}(t)=-\frac{\Omega^2+\beta^2}{\beta}\rho_{vv}(t)+\frac{\Omega^2}{\beta}\rho_{gg}(t).
\end{equation}
Solving Eq. \ref{eq:RHOvv2} using an integrating factor again 
we have
\begin{equation} \label{eq:INTvv}
\begin{split}
\rho_{vv}(t) & = \frac{\Omega^2}{\beta}\int^t_0e^{-\frac{\Omega^2+\beta^2}{\beta}(t-t')}\rho_{gg}(t')d t' \\
 & \approx \frac{\Omega^2}{\Omega^2+\beta^2}\rho_{gg}(t).
\end{split}
\end{equation}
Finally, solving Eq. \ref{eq:RHOve} using the same methodology as above 
we find
\begin{equation} \label{eq:INTve}
\begin{split}
\rho_{ve}(t) & = -\frac{i\Omega}{2}\int^t_0e^{-(\frac{\Gamma_1+\beta}{2}+\gamma)(t-t')}\rho_{ge}(t')d t' \\
 & \approx -i\frac{\Omega}{\beta+\Gamma_1+2\gamma}\rho_{ge}(t).
\end{split}
\end{equation}
Making a change of variables to the saturation parameter $S=\Omega^2/\beta\Gamma_1$ defined in the section on coherent non-resonant driving, we recover the ground and excited state optical Bloch equations for the effective two level system
\begin{equation} \label{eq:2LSvv}
\begin{split}
    \dot{\rho}_{ee}(t) & =-\Gamma_1\rho_{ee}(t)+\beta\frac{\Omega^2}{\Omega^2+\beta^2}\rho_{gg}(t) \\
 & \approx -\Gamma_1\rho_{ee}(t)+S\Gamma_1\rho_{gg}(t),
\end{split}
\end{equation}
\begin{equation}\label{eq:2LSgg}
    \dot{\rho}_{gg}(t)\approx\Gamma_1\rho_{ee}(t)-S\Gamma_1\rho_{gg}(t),
\end{equation}
which holds as long as $\beta\gg \Omega$. We can further manipulate this equality as $\Omega=\sqrt{S\Gamma_1 \beta}$, leading to the constraint $\beta\gg S\Gamma_1$. The final optical Bloch equation to consider is the $\dot{\rho}_{ge}(t)$ contribution. This leads to an interesting pre-factor upon substitution of Eq. \ref{eq:INTve} into Eq. \ref{eq:RHOge}, we find
\begin{equation}\label{eq:RHOgesol}
    \dot{\rho}_{ge}(t)=-\frac{S\Gamma_1\beta}{2(\beta+2\Gamma_2)}\rho_{ge}(t)-\frac{\Gamma_1}{2}\rho_{ge}(t)-\gamma\rho_{ge}(t),
\end{equation}
which for $\beta\gg \Gamma_2$ can be simplified to recover the two-level system optical Bloch equation
\begin{equation}\label{eq:RHOge2LS}
    \dot{\rho}_{ge}(t)=-\frac{S\Gamma_1}{2}\rho_{ge}(t)-\frac{\Gamma_1}{2}\rho_{ge}(t)-\gamma\rho_{ge}(t).
\end{equation}

\subsubsection{Indistinguishability from cw measurement}
Modelling the effective two level system found from the adiabatic elimination using a second-order Born-Markov master equation, we have 
\begin{equation}\label{eq:2LS_ME}
\partial_t \rho(t) = \Gamma_1\mathcal{L}_{\sigma}[\rho(t)]+\Gamma_1 S\mathcal{L}_{\sigma^{\dagger}}[\rho(t)]+2
\gamma \mathcal{L}_{\sigma^{\dagger}\sigma}[\rho(t)].
\end{equation}
The driving in this model is captured by the incoherent dissipator with rate $S\Gamma_1$. Using quantum regression theorem we can explicitly solve the second-order perpendicular and parallel cw correlation functions in Eq. \ref{eq:g2paracw} and \ref{eq:g2perpcw} \cite{QRT}. We do so by solving their constituent parts, firstly finding the excited steady state population ${\expval{\sigma^{\dagger}\sigma}_{ss}}=\lim_{t\to\infty}\mathrm{Tr}_S[\sigma^{\dagger}\sigma (e^{\mathcal{L}t}\rho_S(0))]=\frac{S}{1+S}$, where initially the system is populated is in the ground state. The first order correlation function present in $g^{(2)}_{\parallel_{CW}}(\tau)$ is evaluated to $\lim_{t\to\infty}\expval{\sigma^{\dagger}(t+\tau)\sigma(t)}=\lim_{t\to\infty}\mathrm{Tr}_S[\sigma^{\dagger} e^{\mathcal{L}\tau}\sigma (e^{\mathcal{L}t}\rho_S(0))]=\frac{S}{1+S} e^{-\frac{1}{2}(\Gamma_1(1+S)+2 \gamma)\abs{\tau}}$. Lastly we find, $\lim_{t\to\infty} \expval{\sigma^{\dagger}(t)\sigma^{\dagger}(t+\tau)\sigma(t+\tau)\sigma(t)}=\lim_{t\to\infty}\mathrm{Tr}_S[\sigma^{\dagger} \sigma e^{\mathcal{L}\tau} \sigma (e^{\mathcal{L}t}\rho_S(0))\sigma^{\dagger}]=\frac{S^2}{(1+S)^2}(1-e^{-(1+S)\Gamma_1 \abs{\tau}})$. 
Putting these together we can express the parallel polarisation alignment second order correlation function as
\begin{equation}\label{eq:g2paracwexp}
g_{\parallel_{CW}}^{(2)}(\tau)=\big(1-\frac{1}{2}e^{-(1+S)\Gamma_1 \abs{\tau}}-\frac{1}{2}e^{-(\Gamma_1(1+S)+2\gamma)\abs{\tau}}\big),
\end{equation}
and the perpendicular measurement as
\begin{equation}\label{eq:g2perpcwexp}
g_{\perp_{CW}}^{(2)}(\tau)=\big(1-\frac{1}{2}e^{-(1+S)\Gamma_1 \abs{\tau}}\big).
\end{equation}
This theory can be extended following the work of Ref.~\cite{Rezai2018} to account for visibility (which in this formalism is inherently 1) and extra decoherence effects with no temporal dependence, as shown in the main text. Using these resultant correlation functions we find the numerator in Eq. \ref{eq:ICWcorr} as
\begin{equation}\label{eq:InumCW}
\int^{\infty}_{0}d\tau \lim_{t\to\infty} \abs\Big{ \expval{\sigma^{\dagger}(t+\tau)\sigma(t)}}^2/{\expval{\sigma^{\dagger}\sigma}^2_{ss}}=\frac{1}{\Gamma_1(1+S)+2\Gamma_1}.
\end{equation}
Similarly, for the denominator of Eq. \ref{eq:ICWcorr} we have
\begin{equation}\label{eq:IdemonCW}
\int^{\infty}_{0}d\tau~1-\lim_{t\to\infty} \expval{\sigma^{\dagger}(t)\sigma^{\dagger}(t+\tau)\sigma(t+\tau)\sigma(t)}/{\expval{\sigma^{\dagger}\sigma}^2_{ss}}=\frac{1}{\Gamma_1(1+S)}.
\end{equation}
Substituting these into Eq. \ref{eq:ICWcorr} we find for a two-level system
\begin{equation}\label{eq:ICWana}
\tilde{\mathcal{I}}_{2LS}(S)= \frac{\Gamma_1(1+S)}{\Gamma_1(1+S)+2\gamma}.
\end{equation}
In the limit of $S\rightarrow 0$ we recover the indistinguishability of the system found from the pulsed case $\tilde{\mathcal{I}}_{2LS}(S\rightarrow0)=\frac{\Gamma_1}{\Gamma_1+2\gamma}=\mathcal{I}$. Considering the case of differing driving strengths for the parallel $(S_1)$ and perpendicular $(S_2)$ alignment measurements we find
\begin{equation}\label{eq:IcwS1S2}
\tilde{\mathcal{I}}_{2LS}(S_1,S_2)= \frac{\Gamma_1(1+S_2)}{\Gamma_1(1+S_1)+2\gamma}+\frac{S_2-S_1}{1+S_1}.
\end{equation}

\begin{figure}[t]
\centering
  \includegraphics[scale=0.45]{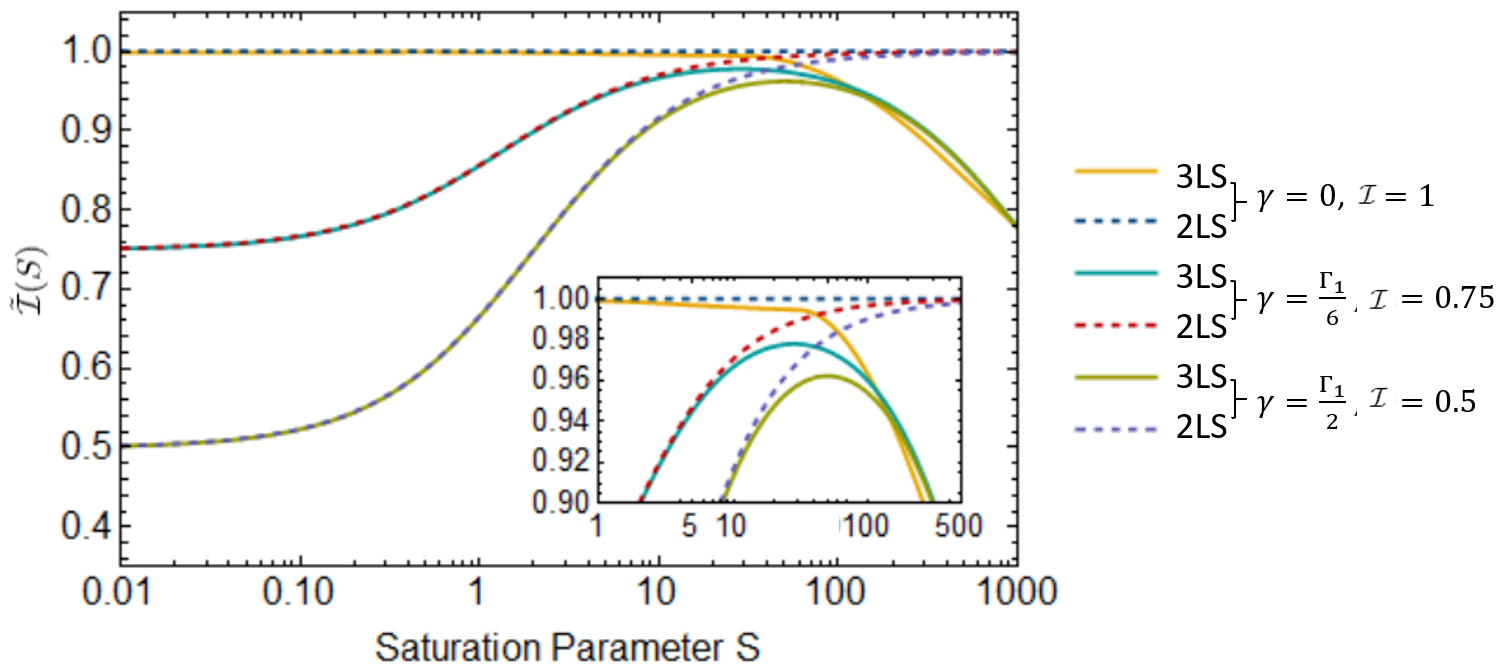}
\caption{Calculation of $\tilde{\mathcal{I}}(S)$ for the coherently driven three level system and the effective incoherently driven two level system for different pure dephasing parameters $\gamma$. Inset shows the region where divergence between the two models occur.}
\label{fig:gamvar}
\end{figure} 

To compare results from the two- and three-level system models see Fig.~\ref{fig:Iresults}(a) and (b), where the parallel and perpendicular second-order correlation functions are compared for extreme $S$ values. It can be seen that for $S=1$ the two models converge as expected, as we are within the limit of $\beta \gg S\Gamma_1$. Setting $S=500$ to capture very strong driving, the coherently driven three-level system cross-correlation functions decrease in width, where $g_{\perp\,CW}^{(2)}(\tau)$ decreases more than  $g_{\parallel\,CW}^{(2)}(\tau)$. To see how this deviation effects the calculation of $\tilde{\mathcal{I}}(S)$ for the two models, see Fig.~\ref{fig:Iresults}(c). It can be seen that for strong driving the effective two-level system model breaks down and our analytical form $\tilde{\mathcal{I}}_{2LS}(S)$ is no longer valid. To investigate for what $S$ this break down occurs at we plot $\tilde{\mathcal{I}}(S)$ for varying $\mathcal{I}$, see Fig~\ref{fig:gamvar}. We find for the case of maximum indistinguishability $\mathcal{I}=1$ the function $\tilde{\mathcal{I}}(S)$ deviates between the two level and three level system models by $\SI{0.5}{\percent}$ at $S=23.3\pm0.1$, for our system parameters. For decreasing $\mathcal{I}$ (as the pure dephasing increases), this break down value of $S$ increases, where for $\mathcal{I}=0.5$ a deviation of $\SI{0.5}{\percent}$ occurs at $S=40.03\pm0.1$. The observed shift in the breakdown of the two-level system model can be explained by looking at Eq.~\ref{eq:RHOgesol}. When the excess pure dephasing becomes non-negligible this acts to suppress the saturation parameter present in this equation.

\begin{figure}[ht!]
\centering
  \includegraphics[scale=0.55]{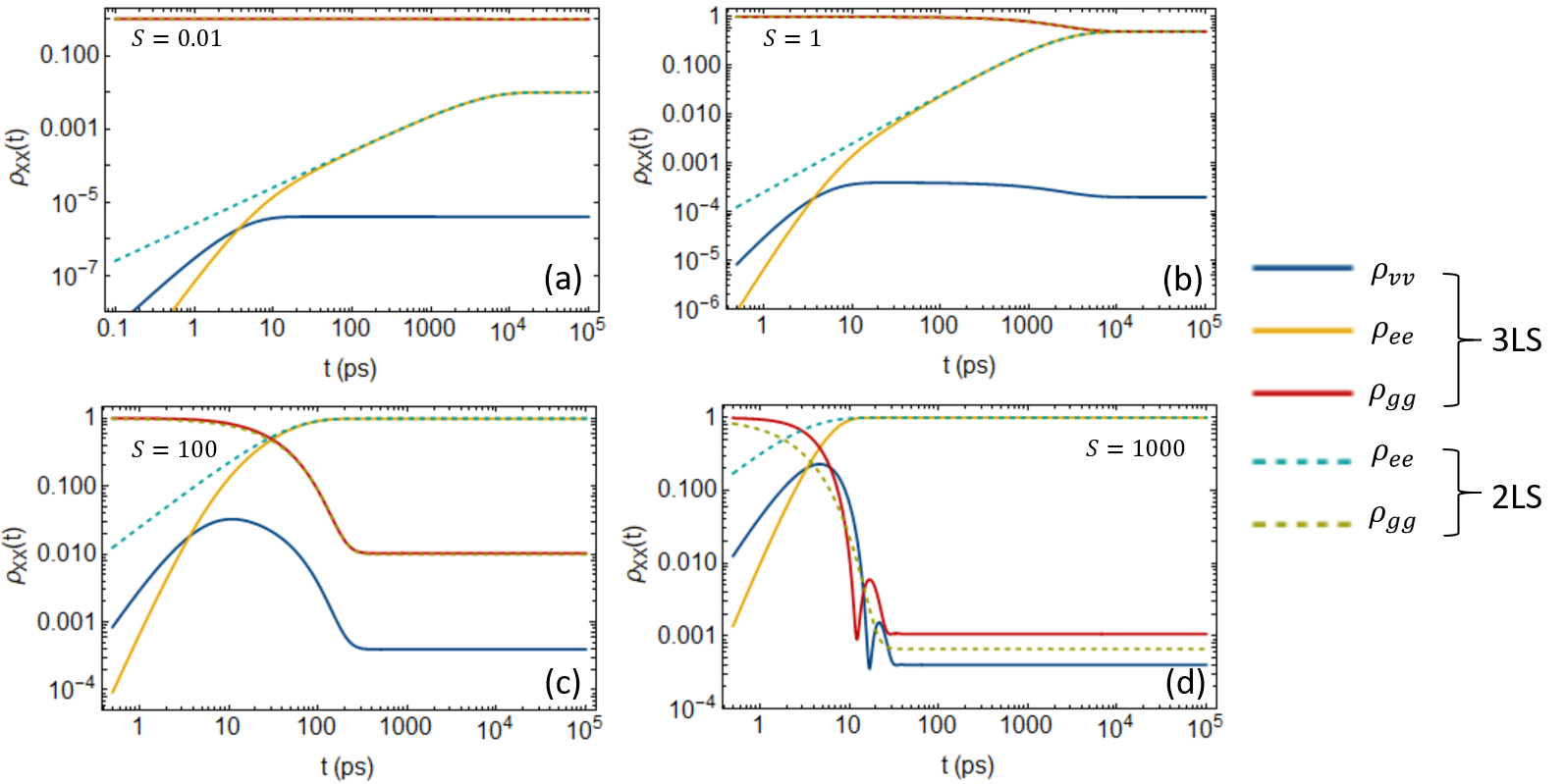}
\caption{Evolution of the states given by $\rho_{XX}(t)=\expval{\rho(t)}{X}$ for both the coherently driven three level system and the effective two level system model for various driving strengths, modified by the saturation parameter $S$. The initial offset for $t<\SI{50}{\pico\second}$ between the two models for the excited state is due to some proportion of the population residing in the vibrational level for the three level system model.}
\label{fig:densityevo}
\end{figure}

The origin for this deviation stems from including the pump level in the system. By driving to the pump level coherently we capture the possibility for coherent exchange between the ground and pump level, which acts to suppress $\tilde{\mathcal{I}}(S)$. To see how the states evolve in the two different models see Fig. \ref{fig:densityevo}, where this coherent exchange can been seen for $S=1000$. From this analysis we can conclude to extract the indistinguishability from a cw measurement using $\tilde{\mathcal{I}}_{2LS}(S)$ in Eq. \ref{eq:ICWana} one must not pump too hard to ensure coherence effects between the ground and pump level can be neglected.

\subsubsection*{Contribution from a sideband}
Until now we have neglected any effect from phonon side bands. This methodology to find indistinguishability can be readily extended to capture the influence of a broad sideband with annihilation operator $b_\mathbf{k}$ for wavevector $\mathbf{k}$, frequency $\omega_\mathbf{k}$ and electron phonon coupling constant $g_\mathbf{k}$. Considering polaron theory we can find the electric field operator by solving the Heisenberg equations of motion to find
$E^{(+)}(t)\propto \sigma(t)B_-(t)$ where
$B_{\pm}=\mathrm{exp}[\pm \sum_\mathbf{k}g_\mathbf{k}(b_\mathbf{k}^{\dagger}-b_\mathbf{k})/\omega_\mathbf{k}]$ is the phonon bath displacement operator \cite{ilesmith2016,Clear2020}. Substituting this into the cross correlation functions $g^{(2)}_{\perp/\parallel_{CW}}(\tau)$ and making the assumption that we can factorise out the photon bath correlation functions due to largely differing time scales results in modification to Eq.~\ref{eq:ICWcorr} by the Debye-Waller factor such that
\begin{equation}\label{eq:ICWSB}
\tilde{\mathcal{I}}(S)=\frac{ \int d\tau \abs{\mathcal{G}(\tau)}^2  \lim_{t\to\infty} \abs\Big{ \expval{\sigma^{\dagger}(t+\tau)\sigma(t)}}^2/{\expval{\sigma^{\dagger}\sigma}^2_{ss}} }{\int d\tau(1-\lim_{t\to\infty} \expval{\sigma^{\dagger}(t)\sigma^{\dagger}(t+\tau)\sigma(t+\tau)\sigma(t)}/{\expval{\sigma^{\dagger}\sigma}^2_{ss}})},
\end{equation}
where we have defined the bath phonon correlation function $\mathcal{G}(\tau)= \expval{B_-(\tau)B_+}$ \cite{Nazir2016}. Using this methodology it is straight forward to extend this model to include contributions from local vibrational modes as well. 

\newpage
\section*{Experimental}
\subsection*{Molecule Characterisation}
\begin{figure}
    \centering
    \includegraphics[scale=0.45]{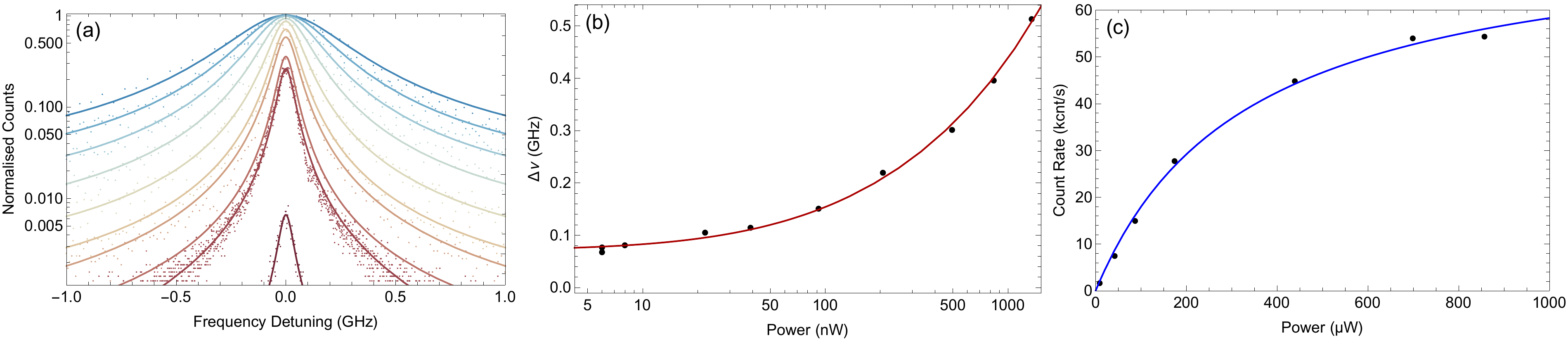}
    \caption{(a) Data is from scanning the laser across the $S_{0,0}\to S_{1,0}$ transition whilst monitoring red-shifted fluorescence at increasing excitation powers. The effect of power broadening with high excitation power is visible. Lines are Lorentzian functions fit to extract the linewidth. (b) Data is linewidth at each power from (a), red line is Eq.~\ref{eq:pb} giving $\Gamma_2= 2 \pi \times \SI{35(4)}{\mega\hertz}$. (c) Data is the count rate from the molecule whilst exciting with blue-detuned \SI{766.67}{\nano\meter} light, blue line is Eq.~\ref{eq:sat}.}
    \label{fig:char}
\end{figure}
To extract the dephasing rate $\Gamma_2$ and population decay rate $\Gamma_1$ a number of characterisation experiments were performed. The dephasing rate is extracted from performing resonant linescans at a range of excitation powers whist monitoring the red-shifted fluorescence, as shown in Fig.~\ref{fig:char}. These are fitted with Lorentzian curves to extract the linewidth \cite{Grandi2016}. The power-broadening relationship is described by
\begin{equation}
    \Delta\nu=\frac{\Gamma_2}{\pi}\sqrt{1+S} \, 
    \label{eq:pb}
\end{equation}
with $S=P/P_{\text{sat}}$, where $P$ is excitation power and $P_{\text{sat}}$ is the power at saturation. This can be used to extract the `zero-power' linewidth, or $2\Gamma_2$ from this data. Figure~\ref{fig:char}(b) shows a plot of this data, with $P_{\text{sat}}=\SI{27\pm3}{\nano\watt}$ and $\Gamma_2=2\pi \times \SI{35(4)}{\mega\hertz}$.

To extract $\Gamma_1$ from the blue-pumped $g^{(2)}(\tau)$ measurement shown in Fig.~2 of the main paper the saturation parameter of the measurement is required. The power is known from measurement of the excitation power, and $P_{\text{sat}}$ is found from a saturation measurement. For the blue excitation a series of 2D spatial scans were performed at increasing power whilst monitoring the fluorescence, as a function of position. These can be fit with 2D-Gaussian curves to extract the maximum count rate at each power \cite{Schofield2018}. This data can then be fit with a saturation curve of the form
\begin{equation}
    R=R_{\infty}\frac{S}{1+S}
    \label{eq:sat}
\end{equation}
where $R$ is the measured count rate and $R_{\infty}$ is the maximum count rate. This is shown in Fig.~\ref{fig:char}(c). We find $P_{\text{sat}}=0.33(3)$\,mW. This can then be used find $\Gamma_1=2 \pi\times\SI{40(2)}{\mega\hertz}$, as described in the main paper.

\subsection*{Filtering}
\begin{figure}
\centering
  \includegraphics[scale=0.7]{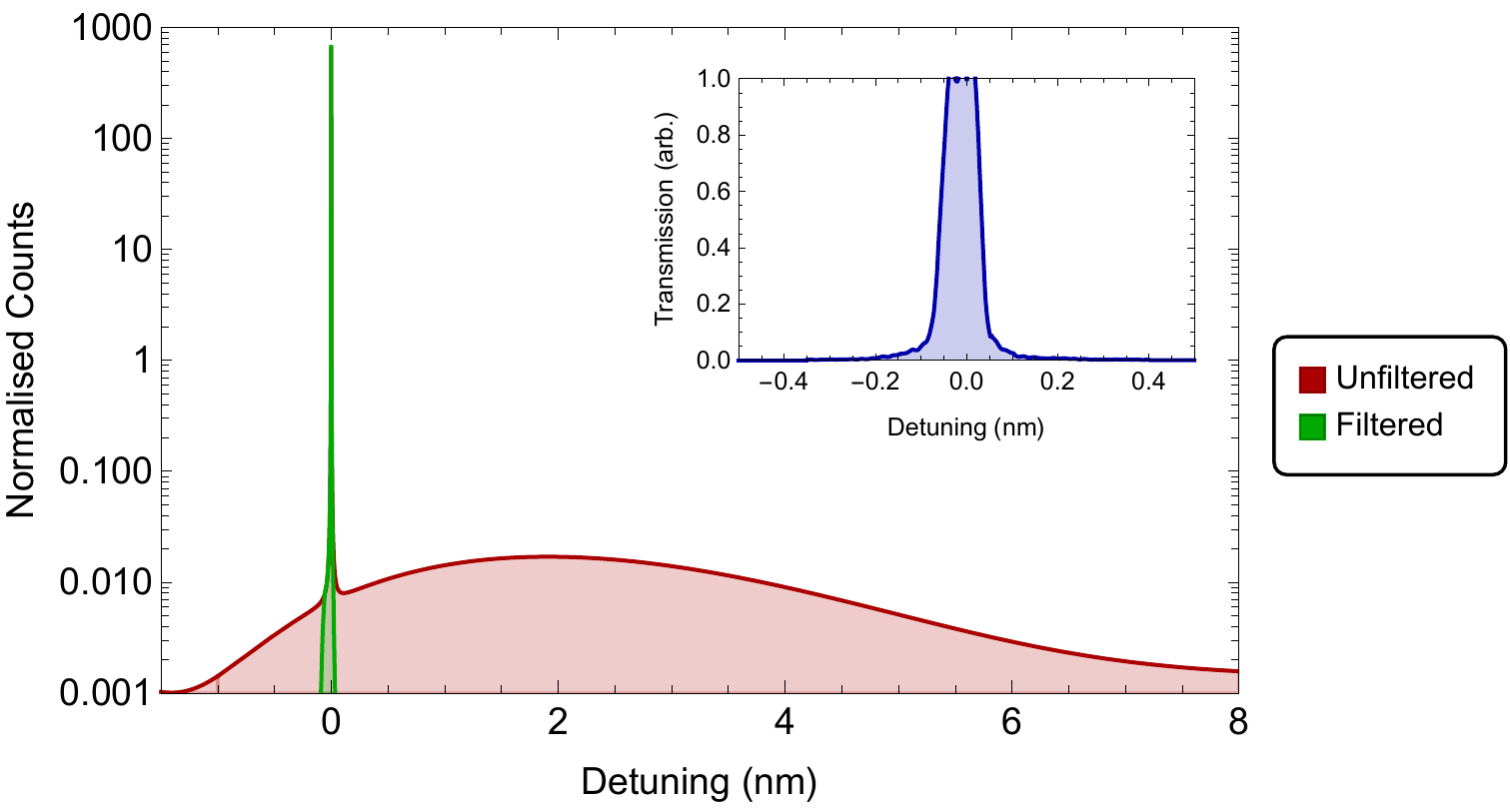}
\caption{A plot showing the unfiltered (red) and filtered (green) spectra after convolution with the filter response function of a \SI{0.15}{\nano\meter} notch reflection filter. The central frequency of the filter can be shifted by tilting the filter, ideally to match the spectrum of the molecule. The arrangement shown, with all of the emission to vibrational modes and nearly all of the sideband removed, has an $\alpha= \SI{99.7}{\percent}$, and a corresponding $\mathcal{I}=\SI{99.3}{\percent}$. Inset: Measured response function of the filter. }
\label{fig:filter}
\end{figure}
In order to remove background and molecular emission on the zero-phonon-line sideband and vibrational transitions, a narrow reflection filter is used before the interferometer. This filter can easily be tuned, by changing angle relative to the emission, to a specific ZPL wavelength. Figure~\ref{fig:filter} shows the measured reflection spectrum of the filter and the effect of the convolution of this with a model DBT spectrum. A spectrum from Ref.~\cite{Clear2020} is recreated using a Lorentzian ZPL and a Poissonian distribution for the phonon sideband. The integrals of these functions are matched to the respective components of the spectrum. Looking at the post-convolution spectrum we find a proportion of coherent ZPL to total collected emission, termed $\alpha$ of $\alpha=\SI{99.7}{\percent}$. For indistinguishability we are interested in the probability of two ZPL photons, giving $\alpha^2=\SI{99.3}{\percent}$ \cite{Clear2020}.

\subsection*{Interferometer Description}
The relative reflection/transmission coefficients of two fibre beam splitters that make up the interferometer have an affect on the relative magnitudes of features in the $g^{(2)}_{\perp/\parallel_{CW}}(\tau)$ measurements. This can be described with the equation
 \begin{align}
g^{(2)}_{\parallel/\perp_{CW}}(\tau) =1 - \frac{\mathcal{V}}{(r_1^2t_0^2+r_0^2t_1^2)(r_0^2r_1^2+t_0^2t_1^2)} \bigg(& r_1^2 t_1^2 (r_0^4 + t_0^4) e^{(-\Gamma_1(1+S)|\tau|)}\nonumber \\ 
 +&\mathcal{M}_{\parallel/\perp} 2 r_0^2 r_1^2 t_0^2 t_1^2 e^{(- (\Gamma_1(1+S)+2\gamma)|\tau|)} \nonumber \\
 +&r_0^2 r_1^4 t_0^2 e^{(-\Gamma_1(1+S)(|\tau - d\tau|))}  \nonumber \\ +& 
 r_0^2 t_1^4 t_0^2 e^{(-\Gamma_1(1+S)(|\tau + d\tau|))} \bigg),
 \label{eq:toro}
    \end{align}
where $\mathcal{V}$ is the anti-bunching visibility, $\Gamma_1$ is the population decay rate, $\gamma$ is the excess dephasing, $d\tau$ is the time delay caused by the delay fibre, and $r_0/t_0$ and $r_1/t_1$ are the reflection and transmission amplitude coefficients for the first and second beam splitters respectively. The mode overlap $\mathcal{M}_{\parallel/\perp}$ is to account for non-perfect polarisation control during each measurement. This is similar to that found in Ref.~\cite{Rezai2018}, but here combined with the driving relationship developed in this work. There are four terms, two corresponding to the anti-bunching (first term) and indistinguishability (second term) contributions to the central feature, and two corresponding to the anti-bunching features at $\pm d\tau$.

Before taking quantum interference measurements the interferometer was independently characterised. We find $t_0=\sqrt{0.501(1)}$, $t_0=\sqrt{0.499(1)}$, $r_1=\sqrt{0.482(1)}$ and $t_1=\sqrt{0.518(1)}$.  The difference in propagation time through the two arms of the interferometer, caused by the long delay fibre in one arm, was \SI{24.75}{\nano\second}, found from the time difference between the side dips in the $g^{(2)}_{\parallel/\perp_{CW}}(\tau)$ measurements. As the interferometer is not actively stabilised there was polarisation drift over the length of a measurement. Monitoring the maximised the output of a PBS on one output of the interferometer we find a \SIrange{2}{4}{\percent} reduction in counts over half an hour, approximately the length of time of a single $g^{(2)}_{\perp/\parallel}(\tau)$ measurement. As $\mathcal{I}$ is based on the difference between the parallel and perpendicular measurements the polarisation drift on both contributes. This gives an approximately \SI{5}{\percent} reduction in $\mathcal{M}$, and therefore in $\mathcal{I}$.

\subsubsection{Pulsed Modelling}
For the case of matched laser repetition rate and interferometer delay Eq.~\ref{eq:toro} can be modified to describe the pulsed measurement with
 \begin{align}
G^{(2)}_{\parallel/\perp_{PUL}}(\tau) =\sum_i \exp(-\Gamma_1(|\tau-id\tau|)) - \frac{\mathcal{V}}{(r_1^2t_0^2+r_0^2t_1^2)(r_0^2r_1^2+t_0^2t_1^2)} \bigg(& r_1^2 t_1^2 (r_0^4 + t_0^4) e^{(-\Gamma_1|\tau|)}\nonumber \\ 
 +&\mathcal{M}_{\parallel/\perp} 2 r_0^2 r_1^2 t_0^2 t_1^2 e^{(- (\Gamma_1+2\gamma)|\tau|)} \nonumber \\
 +&r_0^2 r_1^4 t_0^2 e^{(-\Gamma_1(|\tau - d\tau|))}  \nonumber \\ +& 
 r_0^2 t_1^4 t_0^2 e^{(-\Gamma_1(|\tau + d\tau|))} \bigg),
 \label{eq:toroP}
    \end{align}
with $i$ an integer describing the number of pulses each peak is away from the central $\tau=0,~i=0$ feature. This does not hold for large timing mismatch between the pulse repetition period and time delay from the unbalanced interferometer. Each peak will have contributions from coincidences with different path combinations and effectively an interferometer-parameter weighted sum of these will modify Eq.~\ref{eq:toroP}. More precisely, there are four path combinations for a coincidence count: no delay, no delay; no delay, delay; delay, no delay; delay, delay. Coincidences taking different paths will be offset by the timing mismatch $\pm \Delta\tau$, however the total coincidences per pulse will be unchanged.

From measuring the time difference between photon detections after sending the pulsed laser through a beam splitter we find the pulsed laser has a repetition period of 12.55\,ns corresponding to a repetition rate of 79.7\,MHz. Comparing this to the interferometer delay found for cw measurements above we find a mismatch of \SI{0.4}{\nano\second} compared to twice the laser repetition period. The exponential of the ratio of timing mismatch to the emitter lifetime gives the correction factor used for the pulsed excitation measurements in the main manuscript. For our experimental parameters this offset is smaller than the detector timing uncertainty and as such has no discernible effect after Eq.~\ref{eq:toroP} is convolved with the detector response function. Additionally, as we integrate the data directly the pulse area, or total coincidences per pulse, is the property of interest. 


\subsection*{Indistinguishability Measurements}

\begin{figure}[t!]
\centering
  \includegraphics[scale=0.5]{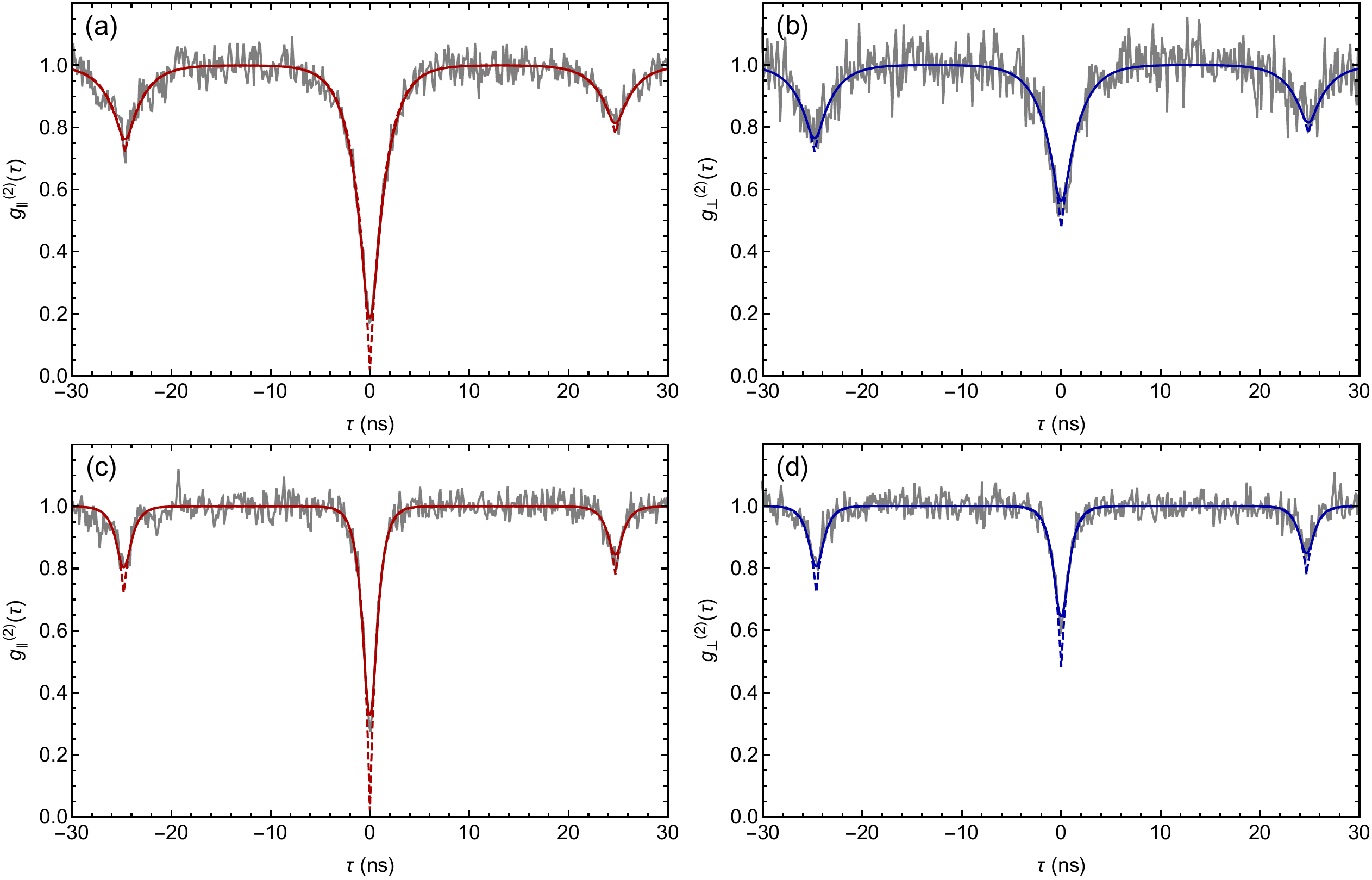}
\caption{(a) Continuous wave (cw) excitation $g^{(2)}_{\parallel_{cw}}(\tau)$ measurement on a single DBT molecule showing anti-bunching and two-photon interference taken at $S=1.3\pm0.1$. Data points are shown with the black line. Overlaid is Eq.~\ref{eq:toro} (dashed red line), and this convolved with the detector response function (solid red line). Parameters are from independent measurements performed on the molecule. The features at $\tau=\pm\SI{25}{\nano\second}$ are due to a combination of anti-bunching and the interferometer delay. (b) A cw excitation $g^{(2)}_{\perp}(\tau)$ measurement on the same DBT molecule. Anti-bunching is visible, however there is negligible interference. As in (a), Eq.~\ref{eq:toro} (dashed blue line) and the convolved form (solid blue line) are overlaid. (c) The same as (a) for $S=4.4\pm0.2$. (d) The same as (b) for $S=4.4\pm0.2$.}
\label{fig:1and4}
\end{figure}

Figure~\ref{fig:1and4}(a) and (b) are presented in the paper as $g^{(2)}_{\perp/\parallel_{cw}}(\tau)$ measurements taken at $S=1.3\pm0.1$. Parameters for the overlaid function are taken from either interferometer characterisation or global fits of the side features. Figure~\ref{fig:1and4}(c) and (d) are the the same measurements performed with $S=4.4\pm0.2$. The narrowing effect of faster driving \cite{Grandi2016} is visible when comparing Fig.~\ref{fig:1and4}(a) and (b) with Fig.~\ref{fig:1and4}(c) and (d). We note that it is also possible to fit the $g^{(2)}_{\parallel_{cw}}(\tau)$ measurement to extract $\Gamma_2$ for cases where $\Gamma_2$ cannot be measured through resonant linescans. For the $S=1.3\pm0.1$ $g^{(2)}_{\parallel_{cw}}(\tau)$ measurement, fitting with Eq.~\ref{eq:toro} for $\gamma$, we find $\Gamma_2=2\pi\times\SI{35(5)}{\mega\hertz}$. 


\bibliography{BIBO}